\documentclass[pre,twocolumn,superscriptaddress,showpacs,aps,floatfix]{revtex4-1}
\setcitestyle{square}
\usepackage[utf8]{inputenc}
\usepackage{graphicx,url, microtype, graphicx, dcolumn, bm, dsfont, csquotes, dsfont, amsmath, amsbsy,amsfonts,appendix, booktabs, amssymb}
\usepackage[dvipsnames]{xcolor}
\usepackage[colorlinks,linkcolor=blue,citecolor=blue]{hyperref}

\usepackage[utf8]{inputenc}
\usepackage[T1]{fontenc}
\usepackage[margin=1in]{geometry}
\usepackage[sort&compress]{natbib}
\usepackage[skins,theorems]{tcolorbox}
\usepackage{soul}
\usetikzlibrary{intersections, angles, fadings, through, positioning,arrows,shapes,automata,petri,positioning,calc}

\newcommand{\bs}[1]{{\boldsymbol{#1}}}

\def\resetul{\setul{0.5ex}{0.3ex}}
\resetul
\def\SOUL@stpreamble{%
    \dimen@\SOUL@ulthickness
    \dimen@i=-1ex
    \advance\dimen@i-1\dimen@
    \edef\SOUL@uldepth{\the\dimen@i}%
    \let\SOUL@ulcolor\SOUL@stcolor
    \SOUL@ulpreamble
}
\setstcolor{red}

\mathchardef\mhyphen="2D
\bibstyle{aps}

\begin{document}
\title{Information-theoretic description of a feedback-control Kuramoto model}
\author{Damian R Sowinski}
\email{Damian.Sowinski@Rochester.EDU}
\affiliation{Department of Physics and Astronomy, University of Rochester}
\author{Adam Frank}
\email{afrank@pas.rochester.edu}
\affiliation{Department of Physics and Astronomy, University of Rochester}
\author{Gourab Ghoshal}
\email{gghoshal@pas.rochester.edu}
\affiliation{Department of Physics and Astronomy, University of Rochester}

\begin{abstract}
{Semantic Information Theory (SIT) offers a new approach to evaluating the information architecture of complex systems.  
In this study we describe the steps required to {\it operationalize} SIT via its application to dynamical problems. 
Our road map has four steps: 
(1) separating the dynamical system into agent-environment sub-systems; 
(2) choosing an appropriate coarse graining and quantifying correlations; 
(3) identifying a measure of viability;
(4) implementing a scrambling protocol and measuring the semantic content.
We apply the road map to a model inspired by the neural dynamics of epileptic seizures whereby an agent (a control process) attempts to maintain an environment (a base process) in a desynchronized state. 
The synchronization dynamics is studied through the well-known Kuramoto model of phase synchronization. 
Our application of SIT to this problem reveals new features of both semantic information and the Kuramoto model.  
For the latter we find articulating the correlational structure for agent and environment (the oscillators), allows us to cast the model from a computational (information theoretic) perspective, where the agent-environment dynamics form a communication channel. 
For the former we find a topological dependence of whether information can be considered semantic or not.
}
\end{abstract}

\maketitle
\section{Introduction}

It is generally accepted that information plays an important role in complex-adaptive, and living systems in particular~\cite{schrodinger1944life}. 
From DNA transcription to signal transduction, information flows and information processing appear to be an essential aspect of living systems~\cite{watson1953molecular,Smith_2007,mattingly2021escherichia,kuppers1990information,adami2004information,uda2020application,gohari2016information,rhee2012application,tkavcik2016information,donaldson2010fitness,rivoire2011value,hazen2007functional}.  
Indeed, the capacity to use information, rather than just be described in terms of it, may be the characteristic which separates life from other physical systems~\cite{sowinski2022consensus, oh2020towards, bowen2022visual}.  

There is, however, an important caveat in making information a central component in developing a {\it physics of life}.  
Information theory, tracing its roots back to Shannon's famous paper~\cite{shannon1948mathematical}, is constructed using purely syntactic measures.  
Information, in this formalism, is defined in terms of the combinatorics of strings composed of letters drawn from an alphabet.
Such a definition purposely ignores the question of the {\it semantic content} of information.  
While highly useful for engineering issues like the design of computational technologies, when it comes to questions associated with the emergence and dynamics of living systems, ignoring issues of meaning filters out the most important ways life uses information~\cite{Schlosser_1998,Mossio_2009}.  
For example, understanding the emergence of agency and autonomy revolve around the capacity of a system to parse streams of information in terms of their {\it valence} for the system i.e. their meaning. 
Indeed, such valence, defined as {\it Viability} has been identified as a critical feature defining the boundary between living and non-living systems~\cite{barandiaran2014norm,egbert2023behaviour}.

While extensive discussions of, and definitions for, semantic information exist, they have mostly focused on issues in philosophy or applications within computer science~\cite{Polani_2001,Thompson_2009,nehaniv_meaningful_2003,barham1996dynamical,deacon2007shannon,corning2007control, gleiser2018we,gleiser2018we,sowinski2016complexity}.  
What has been lacking for a {\it physics of life} is a mathematically precise and scientific applicable theory of semantic information that can inform a wide range of domains ranging from neuroscience, collective microbial behavior, social physics and Origin of Life studies~\cite{Acebron_2005, Garcia_2021, Cooper_2013, Mimar_2019, Cohen_1977, Rooney_2006,Xavier_2020, Craig_2014,Mimar:2021si}. 

In~\cite{kolchinsky2018semantic} (KW18) such a semantic theory of information (SIT) was proposed. 
Their formulation uses the state spaces and probability distributions for an agent $A$ and its environment $E$ to characterize the mutual information between the two, while the persistence of $A$ (its ability to maintain a desired state) is measured through a real-valued viability function $\mathcal V$.
The concept of meaning here is thus taken in the most basic sense of being related to an agent's continued existence.
By running {\it intervened} versions of the system dynamics in which some fraction of the mutual information between agent and environment is scrambled, a formal working definition of the semantic information was characterized in terms of the response of the viability function to such interventions.  
Importantly, the viability is determined by the inherent coupled dynamics of the system and the environment~\cite{kolchinsky2018semantic,rovelli_meaning_2018}, rather than through exogenous utility, cost, error, or loss functions (as is sometimes done when studying the value of information in statistics or in engineering applications~\cite{sowinski2021poroelasticity,stratonovich_theory_2020,shannon1959coding,hazen2007functional,sowinski2021poroelasticity}).

The KW18 formalism appears to offer an attractive path forward for the development of a working theory of semantic information.  
The difficulty, however, lies in the computational cost of applying the formalism to real world applications. 
Calculating the information theoretic measures require a precise elucidation of the state spaces, and the computation of the joint probability distributions between the agent and the environment. 
Unfortunately such state spaces are often of such high dimensionality that the formalism is computationally intractable. 
Thus if SIT is to live up to its promise, work must be done to find {\it effective} measures for the quantities required in its formalism~\cite{pemartin2024shortcuts}.  
This includes best practices for definitions of viability, proper mechanisms for scrambling shared information in the intervened trajectories and, finally, practical methods of dimensional reduction (i.e. coarse-graining) for calculations of information theoretic metrics.

Recently, an attempt to develop such effective versions of the KW18 formalism was presented in~\cite{Sowinski:2023vf}.
The authors studied a forager model, a minimal model that captures attributes of living systems, namely  exploration and resource consumption~\cite{benichou2014depletion, bhat2017does, benichou2016role, bhat2022smart}. 
The viability was defined as the expected lifetime of the forager and the scrambling was done by noising a sensor that the forager used to detect resources. 
They applied and extended the work of KW18 and demonstrated how semantic information measures provide new insights into forager dynamics. 
More importantly, they provided a first cut at developing effective measures for application of the SIT formalism to real problems.  

An important result from~\cite{Sowinski:2023vf} was the identification of a plateau in the viability curve as a function of scrambling correlations between agent and environment.  
This plateau occurs at a {\it viability threshold} that is well below channel capacity. 
In other words a subset of correlations between agent and environment has no effect on the forager's lifetime (the information was not semantic). 
Once the threshold is breached the ability for the forager to survive decreases monotonically with increased scrambling (noise in the sensor).
Thus applying the ideas of SIT results in an identification of features in the viability curve that independent of the particular foraging strategy used.  
Whether such semantic saturation is a generic feature of living systems is far from settled, but the results make it clear that finding ways to apply the fundamental principle of SIT results in novel ways to analyze systems from an informational perspective.  

In this paper we report on the next step in efforts to operationalize SIT and explore {\it effective} versions of the theory.
To carry out that task we explore the biophysically motivated problem of a control process (CP) acting on a base process (BP) to maintain overall system viability. 
We analyze a model of a network of coupled oscillators which tend towards synchronization (the base process) and an intervening agent (the control process) which uses the information on the network to limit synchronization. 
Our setting is loosely inspired by the neurodynamics of epileptic attacks~\cite{Zheng:2014qp,Uhlhaas_2006,Siegel:2012ta,Ghoshal:2016xw} where brain pathologies occur due to large scale synchronization of neurons~\cite{Penfield_1954}.
We study the synchronization dynamics through the well-known Kuramoto model of phase synchronization~\cite{kuramoto1975self}, which while applicable to a wide range of settings, has also been used to describe complex brain wave activity: resting state functional connectivity~\cite{cabral2011role}, synchronization due to time-delayed signal transmission~\cite{petkoski2019transmission} and the effects of myelination~\cite{petkoski2023white}.

Additionally, in versions of the Kuramoto model with non-local coupling~\cite{Abrams_2004} or symmetry-breaking~\cite{Zhang_2021} there exist chimera states, where identically coupled oscillators coexist in synchronized and desynchronized states. 
Such states have been observed in both brain networks and the resuscitation of cardiac cells~\cite{Omelchenko_2018}. 
Thus desynchronization is important in many biological contexts. 
Consequently, the intervening agent here acts as preventing the system from reaching (a potentially) pathological synchronized state. 
Our goal in this paper, as in~\cite{Sowinski:2023vf}, is to both explore the application of the KW18 formalism and use SIT to shed new light on a previously well-studied problem, phase-(de)synchronization.

In what is to follow, in Sec.~\ref{sec: Roadmap} we explicitly unpack the general steps required for the use of the SIT formalism. 
In Sec.~\ref{sec: DSD} we introduce the feedback-control model on the coupled oscillators. 
We demonstrate that introducing the agent leads to a stochastic version of the original Kuramoto model, with oscillators coupled to a heat-bath mediated by their degree $k_i$, and a deformed distribution of frequencies $g^A(\omega)$. 
Following this in Sec.~\ref{sec: CG} we recast the model in information theoretic terms in which we show that the base process is a stochastic computation, and the control process is akin to analysing a communication channel between the agent and its environment. 
The viability function is defined in Sec.~\ref{sec: vb} and in Sec.~\ref{sec: spsc} we analyze the semantic content of the correlations between agent/environment demonstrating that for a certain topology, there exists a semantic threshold, whereas for other connectivity patterns, there is no threshold, indicating that every bit of information in the communication channel is crucial to the viability of the system.
Finally in Sec.~\ref{sec: disc} we end with a discussion of the implications of our findings. Three Appendixes clarify technical aspects of our model and the approximations used.

\section{Preliminaries}\label{sec: Roadmap}
Here we provide a set of general set of steps for applying semantic information theory to any framework. 
\begin{itemize}
    \item {\bf Dynamical System Decomposition.} 
    Construct a {\it dynamical system} $(S,\mathcal{D})$, where $S$ is the system state space and $\mathcal D:S\times T\rightarrow S$ is the dynamics on the state space for all times $t\in T$.
    The dynamics can be deterministic, stochastic, or a combination of the two; if they are continuous then $T\simeq \mathds{R}$ or some subinterval of the reals, and if they are discrete then $T\simeq\mathds{Z}$, or some subset of the integers.
    The system must be decomposable into an {\it agent} and {\it environment}, $S=A\times E$, meaning that any state of the system can be written $s=(a,e)$ where $a=(a_1,a_2,\dots)\in A$ and $e=(e_1,e_2,\dots)\in E$ are, respectively, the agent and environment dynamical degrees of freedom (DoF).
    We write $s(t) = (e(t),a(t))$ to represent the system evolution through state space, and refer to the specific form of $\mathcal D$ responsible for $a(t)$ as the agent's {\it behavior}.
    $S$ needs to accommodate a (not necessarily unique) ground state for the agent which acts as an attractor for the system dynamics and represents a {\it dead} agent and an evolving {\it agent-free} environment.

    \item {\bf Coarse Grain and Quantify Correlations.}
    Correlations between agent and environment DoF are manifest in the information architecture of the system.
    Information is quantified by a corpus of measures including mutual information, $I(A:E)$~\cite{Cover_thomas}, cross entropy $H(A:E)$~\cite{Chen_2022}, and transfer entropy $\mathcal T_{E\rightarrow A}$~\cite{Schreiber_2000}.
    In practice these quantities may not be calculable or efficiently computable in full, so a bit of finesse is required at this step to find the right information measure to do the job \cite{ibanez2024heating, tejero2024asymmetries}.
    A combination of prior knowledge and intuition must be used to coarse grain the system into DoF that lend themselves to the desired analytical or computational techniques, while capturing the minimal dynamics of $S$ required to describe an agent attempting to survive in and co-evolve with its environment.
    In particular, the agent-free attractor must be preserved by the coarse-graining.
    For example, it would be intractable to describe a bacterium's environment by listing the position and momentum of every atom surrounding it; a simpler description using a coarse-grained glucose density field might suffice.
    Describing the full biochemistry of the bacterium using every molecular DoF is intractable, but a coarse grained smaller set of DoF such as {\it center of mass position}, {\it orientation}, {\it stored ATP density}, and {\it locomotive modality} may not be.
    Note how this coarse-graining preserves the ground state: in the limit when the bacterium's stored ATP supply vanishes it is effectively dead. 
    
    \item {\bf Identify Viability Measure $\mathcal V$.}
    Since the ground state remains an attractor for the system dynamics under coarse-graining, the {\it distance} to it endogenously defines the {\it viability} of an agent.
    {\it Distance} here can mean several things: 
    If the attractor is stable, then an agent trajectory will eventually end up on it, i.e. the agent will die. 
    In this case the expected lifetime of the agent is a proper measure of distance, and hence viability.
    The example of the bacterium falls in this case; the expected first crossing time to reach zero ATP supply is a viability measure.
    If the attractor is unstable, then the coarse-graining from the prior step must be able to identify an order parameter representing the {\it efficacy} of the agent at maintaining  the agent-free out-of-equilibrium environment.
    The efficacy vanishes in an agent-free environment, and takes some non-zero value once the agent is introduced. 
    The expected efficacy is then a proper measure of distance, and hence viability.
    The viability of the agent measured in either case is referred to as the {\it actual viability}.
       
    \item {\bf Scrambling Protocol and measuring semantic Content.}
    The goal of SIT is to identify how the correlations and viability defined in the last two steps relate to one another.
    Specifically, to find which correlations matter to the agent.
    This is done by implementing a scrambling protocol on the correlations, and examining the resultant change in viability due to the impact the scrambled correlations have on the dynamics of the system.
    This was expounded on in KW18 using {\it interventions} to generate counterfactual histories, and requires a search over all possible partitions of the system degrees of freedom.
    Though rigorously defined, the procedure becomes unmanageable for systems with more than several hundred states.
    Rather than going through all partitions, the coarse-graining, if done correctly, identifies which coarse-grained DoF should be investigated and scrambled.
    In the example mentioned earlier, the bacterium's orientation and whether it is tumbling or rolling are correlated to gradients and magnitudes of the glucose density.
    A scrambling protocol could be introduced at the level of the dynamics by intervening on the bacterium's ability to sense either a magnitude or a gradient, or on the conditions that determine when the bacterium should switch it's locomotive modality.
    The protocol should be tunable, in the sense that the actual viability is a limiting case.
    To contrast with the actual viability, the {\it scrambled viability} is measured as the protocol is tuned higher. 
    The point at which the scrambled viability diverges from the actual viability is the {\it semantic threshold}~\cite{Sowinski:2023vf}.

    \end{itemize}
%
\section{Kuramoto Model with Feedback-Control}\label{sec:km_fc}
Kuramoto's famous coupled oscillator model has been well-studied since it's introduction nearly half a century ago \cite{kuramoto1975self}. 
From the synchronization of fireflies, cardiac pacemakers, laser, power transmission relay networks among others, it has proven a rich starting point for modelling systems in which synchronization is prominent~\cite{mirollo1990synchronization,strogatz2005crowd,moiseff2010firefly,li2017optimizing}. 
Above a critical coupling strength, an extensive fraction of oscillators achieve an equilibrium synchronized state, while the desynchronized states are non-equilibrium transients in the dynamics. 
As mentioned in the introduction, the phenomenon of \emph{desynchronization} has biological relevance in terms of pathological states in the human brain and in cardiac arrhythmia. 
Inspired by this, we develop a feedback-control model of an agent acting on a population of coupled oscillators that attempts to \emph{prevent synchronization}. 
We then apply the SIT framework to this idealized model and develop a new information theoretic perspective on coupled oscillators. 
In our formulation, the control process (an agent) seeks to drive the base process (the Kuramoto model) to a desired far-from-equilibrium set of states. 
Remaining far-from-equilibrium provides a definition of viability i.e. it motivates the teleological dynamics of the agent (the control process) seeking to keep the system from complete synchronization. 
Viewing the system through the lens of semantic information the agent can be interpreted as a repair process acting on a malfunctioning background.  

\subsection {Dynamical System Decomposition}\label{sec: DSD}
\subsubsection{Environment Description}\label{sec: 1A1}
The environment is a population of $N$ oscillators $(\theta_1 \dots \theta_N)$, $\forall i, \theta_i\in(-\pi,\pi]$ placed on the nodes of an undirected graph $G$ with connections parameterized by the adjacency matrix ${\bs A}$ whose elements are $A_{ij}=1$ if nodes $i$ and $j$ are connected, and $0$ otherwise. 
In what is to follow, we consider the fully-connected, Poisson and scale-free degree distributions generated via the configuration model~\cite{Newman_nets}.
The oscillators are coupled to one another with a coupling $\sigma$, which we take to be a function of time.
Each oscillator has a natural frequency, $\omega_i$, drawn from a distribution $g(\omega)$.
Without loss of generality we set $\langle\omega\rangle=0$. 
The variance, or more precisely the root mean square frequency $\omega_\text{rms}=\langle\omega^2\rangle^{1/2}$, is used to rescale time, $\tilde t = \omega_\text{rms} t$. Dropping the tildes, the dynamics of the dimensionless environment are the Kuramoto equations of motion, thus,
\begin{equation}\label{eq: Kuramoto dynamics}
    \frac{d\theta_i}{dt} = \omega_i + \sigma \sum_{j=1}^NA_{ij}\sin(\theta_j-\theta_i),
\end{equation}
with $g(\omega)$ now having vanishing mean and unit variance.
Note, that the original variant of the model can be recovered by setting $A_{ij}=1$ (resulting in a fully-connected graph), and identifying $\sigma = K/N$, with $K$ being the unscaled coupling strength.

\subsubsection{Agent Description}\label{sec: 1A2}
\begin{figure}[h!]
\centering
    \includegraphics[width=\columnwidth]{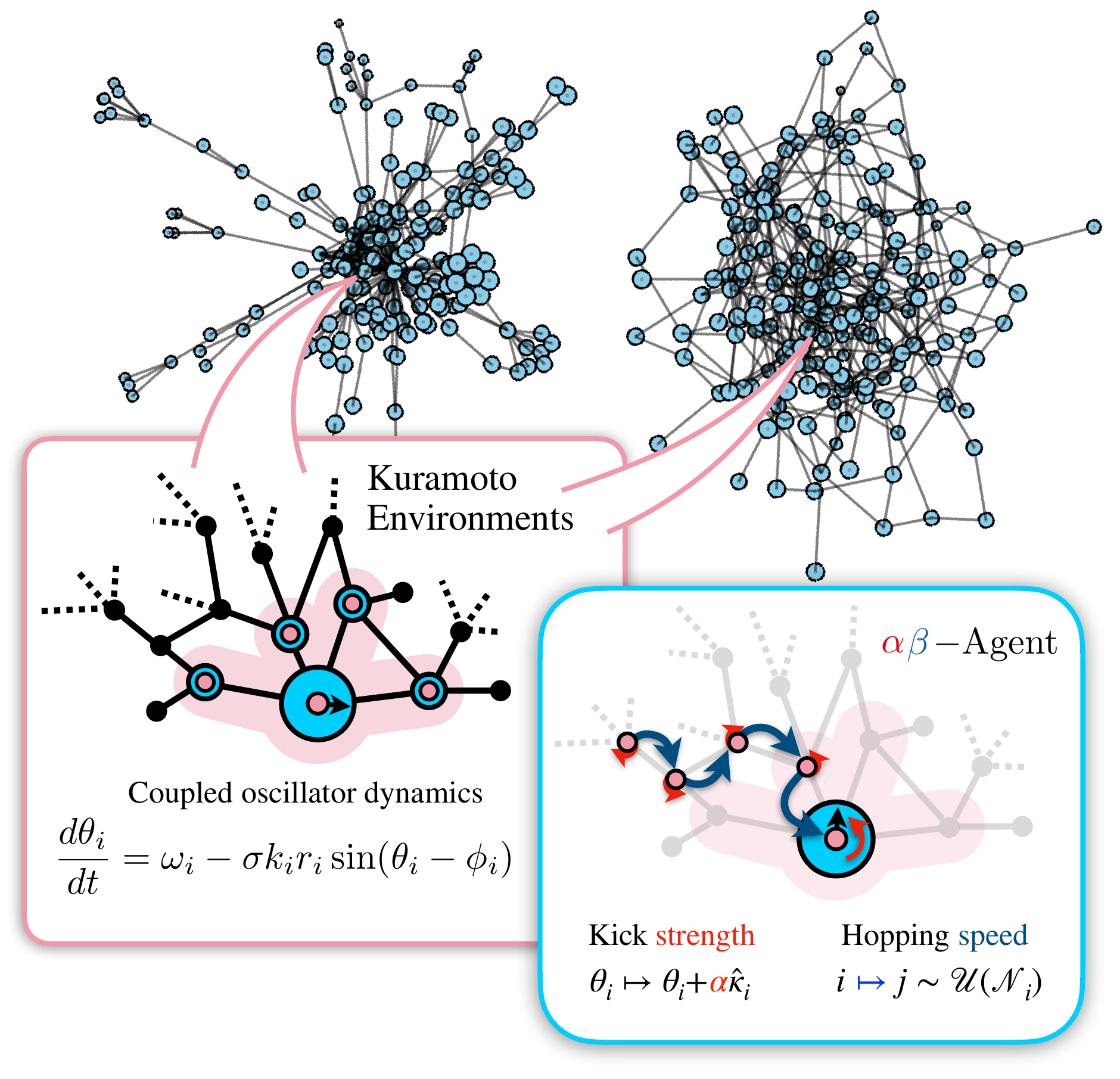}
    \caption{
    Oscillators $\{\theta_i\}$ with natural frequencies $\{\omega_i\}$ are placed on the nodes of different types of networks; shown here power law (upper left) and Poisson graphs (upper right).
    Kuramoto dynamics are imposed resulting in Kuramoto environments. 
    At each node (lower left) there is a local order parameter and local mean field, $\{r_i,\phi_i\}$. 
    Together, these dynamical quantities define the local mean field coupling for each oscillator. 
    Control agents are placed in the Kuramoto environments, moving uniformly at random along edges, and perturbing the nodes based off whether the local mean field estimate is running ahead or behind the oscillator, $\hat{\kappa}_i=\text{sign}(\phi_i-\theta_i)$. 
    Here $\alpha$ describes the strength of the perturbation that the agent applies, and $\beta$ the time-scale or speed of the perturbation.}
    \label{fig:Overview}
\end{figure}

The agent is a localized, mobile perturbation whose purpose is to maintain a global desynchronized state.
The agent's degrees of freedom are $a=(i,\kappa_i )$, where $i$ is the index of the node it is at, and $\kappa_i\in\{\pm1\}$ is the direction of a perturbation of {\it strength} $\alpha$ it applies to the oscillator at $i$. 
After perturbing, the agent \emph{jumps} to an adjacent node uniformly at random; the entire jump-perturbation action takes a time much shorter than the oscillator timescale, $\tau_J \ll 1$, but not so fast that it can perturb more than a small fraction of the network on that same timescale, $\tau_J\gg 1/N$.
This gives a natural way to parameterize the {\it speed} of the agent as $\beta = (N\tau_J)^{- 1}\ll 1$.
As we will show, the single parameter $\alpha\beta$, fully describes the behavior of an agent. 
A schematic of the process is shown in Fig.~\ref{fig:Overview}.

\subsubsection{Dynamics}\label{sec: dynamics}
We begin with the dynamics of the environment. A global order parameter defined in terms of the synchronization $r$, and mean field $\phi$ is constructed thus,
\begin{align}
    r e^{i\phi}=\frac{1}{N}\sum_{j = 1}^N e^{i\theta_j}.
\end{align}
Analogously, local order parameters and mean fields are introduced via
\begin{align}
    r_i e^{i\phi_i}=\frac{1}{k_i}\sum_{j = 1}^NA_{ij}e^{i\theta_j},
\end{align}
where $k_i = \sum_j A_{ij}$ is the degree of node $i$.
In the fully-connected Kuramoto model Eq.\eqref{eq: Kuramoto dynamics} describes a set of oscillators that interact solely through their mutual coupling to the mean field,
\begin{align}\label{eq: Environment EoM 1}
    \frac{d\theta_i}{dt} = \omega_i - Kr\sin(\theta_i-\phi),
\end{align}
with $K$ described earlier.
For arbitrary network topologies 
Eq.\eqref{eq: Kuramoto dynamics} can be written in an analogous way using the local order parameters~\cite{Kuramoto_nets},
\begin{align}\label{eq: Environment EoM 2}
    \frac{d\theta_i}{dt}=\omega_i-\sigma k_ir_i\sin(\theta_i-\phi_i).
\end{align}

In the thermodynamic limit, $N\rightarrow\infty$, of the fully-connected model, above a critical coupling ($K_c = 2/\pi g(0)$) oscillators begin to synchronize, $r>0$.
Analysis of mean field formulations for more complex topologies suggest that $\sigma_c=K_c\langle k\rangle/\langle k^2\rangle$, where $\langle k^n\rangle$ is the $n^\text{th}$ moment of the degree distribution of $G$.
Numerical results suggest a more complicated relationship with topology, while confirming that $\sigma_c\propto K_c$.  
In our model the coupling is driven exogenously, that is perturbations to the system from outside of itself can drive $\sigma$ to values above and below $\sigma_c$ (for instance exposure to stroboscopic lights triggering an epileptic fit, or magnetic fields altering cardiac rhythm). 
Of interest to us will be when the system transitions through the critical coupling from below, shifting the system equilibrium to synchronized states.
As the environment slides towards its new equilibrium, the agent as a control process, seeks to maintain a desynchronized state.  
Thus it must tune its actions towards that goal, while only using local information, that is knowledge gained only at the location $i$ (the node) and its immediate neighborhood.

The agent at node $i$ has access to the phase of the oscillator $\theta_i$. 
It estimates the mean field due to the neighboring oscillators, $\hat\phi_i$, and compares it to $\theta_i$. 
If $\theta_i > \hat\phi_i$, the agent perturbs the oscillator by changing its phase positively, whereas if $\theta_i \leq \hat\phi_i$ it perturbs it in the opposite direction. 
More precisely we have $ \theta_i \mapsto \theta_i + \alpha \kappa_i$, where $\alpha$ is the strength of the agent's perturbation and
\begin{equation}
    \label{eq: Agent perturbation}
    \kappa_i = 
    \begin{cases}
        +1 & \text{if }\theta_i>\hat\phi_i\\
        -1 & \text{otherwise}
    \end{cases}
\end{equation}

\subsubsection{Agent-effect on synchronization dynamics} \label{sec: agent-effect}
\begin{figure}[t!]
    \includegraphics[width=0.9\columnwidth]{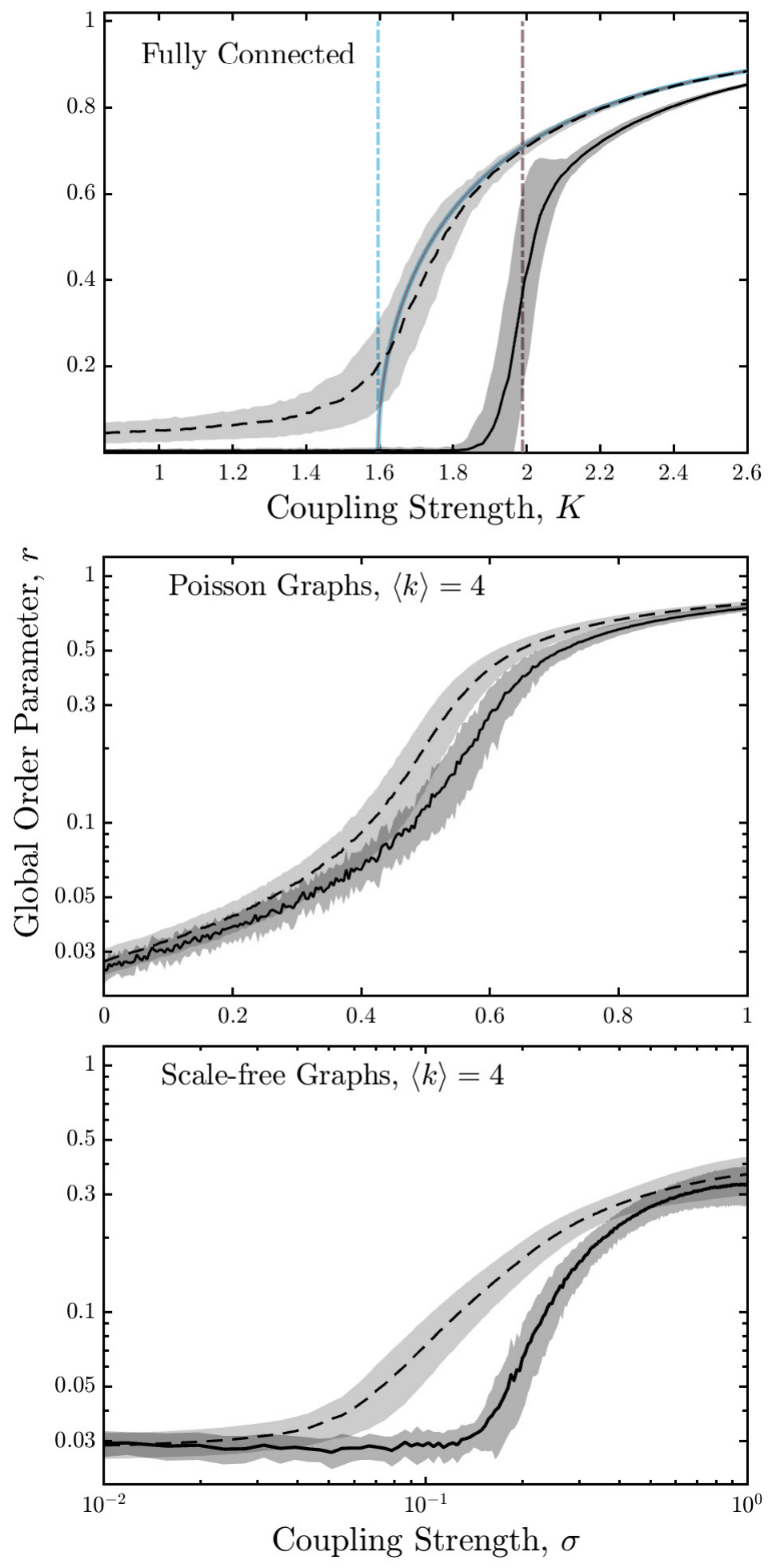}
    \caption{The expected global order parameter for fully-connected (top), Poisson (middle), and scale-free (bottom) graphs ($N = 10^3$) as a function of the coupling strength.
    For the fully connected version, shown as a vertical dashed blue line is the critical coupling $K_c$ in the absence of an agent, and in dashed vertical pink line, the the approximation of the critical coupling in the presence of an agent, $K^A_c$.
    All other networks have an average degree $\langle k \rangle  = 4$. The dashed black curves represent a pure Kuramoto environment, while the solid black curves represent incorporation of the intervening agent. 
    In both cases the shaded regions represents one standard deviation fluctuations from an ensemble computation. 
    Note that the vertical axis scales linearly in the top plot, and logarithmically in the bottom two plots, while the horizontal axis is linear for the top two plots and logarithmic on the bottom one.}
    \label{fig: Order Parameter}
\end{figure}

We simulate an ensemble of agents with $\alpha = 0.5$ and $\beta = 0.16$, first on the fully-connected version of the model, and then on two types of graphs, with the same average degree $\langle k \rangle$ but increasing degree-variance $\langle k^2 \rangle - \langle k \rangle ^2$. 
(See Appendix \ref{app: simulation methodology} for details of the simulation.)
In Figure ~\ref{fig: Order Parameter} we plot the global order parameter $\mathds{E}[r]$, as a function of the coupling strength $\sigma$. 
The dashed curves represent the Kuramoto environment without the agent, whereas the solid curves represent agent incorporation.  
In all cases, we see the effect of the agent is to delay the onset of synchronization, with the most dramatic suppression seen in fully-connected and scale-free graphs.
Simulations varying $\alpha, \beta$ exhibit identical behavior between an $(\alpha,\beta)$ agent and a $(\alpha',\beta')$ agent if $\alpha\beta=\alpha'\beta'$.

To get an analytical handle on why this degeneracy occurs, we incorporate the discrete agent dynamics, Eq. \eqref{eq: Agent perturbation}, into the continuous dynamics of the environment, Eq.\eqref{eq: Kuramoto dynamics}.
The specifics are in Appendix \ref{app: continuum approximation}, but the end result is the stochastic differential equation,
\begin{align}
    \label{eq: Langevin EoM}
    \frac{d\theta_i}{dt}=\omega_i& -\sigma k_ir_i \sin(\theta_i-\phi_i) \nonumber\\
    &+\alpha\beta\frac{k_i}{\langle k\rangle} \kappa_i+\sqrt{\alpha\beta  \frac{k_i}{\langle k\rangle}}\frac{dW_i}{dt},
\end{align}
where $W_i$ is a Wiener process. 
Incorporation of an agent leads to the addition of two dynamical terms in Eq.~\eqref{eq: Kuramoto dynamics} parameterized by the perturbation strength and timescale $\alpha$ and $\beta$, respectively. 
Both parameters enter in solely through their product, providing an explanation for the observed degeneracy.

The two additional terms in Eq.~\eqref{eq: Langevin EoM} have interesting physical interpretations.
The second one plays the role of a heat bath as in other variants of the Stochastic Kuramoto model~\cite{StochasticKM_1, StochasticKM_2,sakaguchi1986soluble,sakaguchi1988phase}.
The key difference here is that the strength of the fluctuations at location $i$ is modulated by the degree of the node $k_i$. 
Instead of a single heat bath, there are several at temperatures quantized in units of $1/\langle k\rangle $.
Nodes with higher degree are coupled to hotter baths, while those with lesser degree to colder ones.

What about the first additional term?
For an oscillator to synchronize, Eq. \eqref{eq: Environment EoM 2} implies it's natural frequency satisfies $|\omega_i|\le \sigma k_ir_i$. 
When $\omega_i$ is positive, the oscillator will do so ahead of the mean phase, while for negative $\omega_i$ it will do so behind the mean phase; put simply $\omega_i$ has the same sign as $\kappa_i$. 
Therefore the term effectively shifts the natural frequencies of synced oscillators to larger values,
\begin{align}\label{eq: frequency shift}
    \omega_i\mapsto \tilde\omega_i= 
    \begin{cases} 
      \omega_i + \alpha\beta\frac{k_i}{\langle k\rangle} & \omega_i>0, \\
      \omega_i - \alpha\beta\frac{k_i}{\langle k\rangle} & \omega_i<0. \\
   \end{cases}
\end{align}
This shift changes the synchronization criterion to $|\omega_i|<\sigma k_i r_i-\alpha\beta k_i/\langle k\rangle$.
Oscillators with positive natural frequencies between $\sigma k_ir_i-\alpha\beta k_i/\langle k\rangle < \omega_i<\sigma k_i r_i$ or negative natural frequencies between $-\sigma k_ir_i < \omega_i<-\sigma k_i r_i+\alpha\beta k_i/\langle k\rangle$ that were synchronized in the absence of an agent will desynchronize in the presence of one.
Consequently the local order parameters, $r_i$, decreases, causing a cascade of oscillators inside the old synchronization range to follow suit until a new equilibrium is reached. 

The combined effect of the two agent terms in Eq. ~\eqref{eq: Langevin EoM} can best be understood qualitatively via their effect on the frequency distribution, $g(\omega)$.
The shift produces a new distribution $\tilde g(\tilde \omega)=g(\omega)$, which vanishes for $|\tilde\omega_i|<\alpha\beta k_i/\langle k\rangle$
In the thermodynamic limit, for a random node in the network $\omega$ is a random variable drawn from this distribution. 
The heat bath introduces a fluctuation to the frequency, another random variable, $\delta\omega$.
The distribution of $\omega+\delta\omega$, the sum of two random variables, is the convolution of their distributions, so the end result is the effective distribution $g^A(\omega) = (\tilde g*h_{\alpha\beta})(\omega)$, where $h_{\alpha\beta}(\delta\omega)$ is a Gaussian with vanishing mean and variance $\alpha\beta$~\cite{widder2015convolution}.
Both operations increase the variance, so $g^A$ is a stretched version of $g$ with a suppressed peak.
Since the critical point depends inversely on this value, this will necessarily increase the critical coupling.

\subsection{Coarse-graining and quantifying correlations.} \label{sec: CG}
The full state space of our system is $S=(-\pi,\pi]^N\times\{1,2,\dots,N\}\times\{\pm 1\}$.
The last of these describe the agent's action $\kappa_i$, based on their observation of the local mean field $\hat\phi_i$.
Combining this with $\theta_i$, the agent determines whether oscillator $i$ is running ahead or behind the mean field of its neighbors.
For the agent, the state space of the environment is coarse grained to a string of bits $\bs{b}$. 
The $i^\text{th}$ bit $b_i$ is $1$ if the oscillator at the $i^{th}$ node is running ahead of its mean field, or $0$ otherwise.
The action of the agent depends on the outcome of its measurement, and the maximal mutual information between the environment and the agent's action degree of freedom is one bit per jump.

As the oscillators evolve, the bits corresponding to synchronized oscillators remain fixed, whereas bits corresponding to unsynchronized oscillators flip pseudo-randomly.
In this light, the Kuramoto model with feedback control can be reduced to two coupled stochastic computations.
The timescale between flips of the computing bits from Eq.\eqref{eq: Environment EoM 2}, under a steady state assumption on both $r_i$ and $\phi_i$, is half an oscillators period
\begin{align}\label{eq: timescale}
    \tau_f &= \frac{1}{2}\int_0^T\!\!\!\! dt = \frac{1}{2}\int_{-\pi}^\pi \!\!\! d\theta \ \frac{1}{\dot\theta}\nonumber\\
    &=\frac{\pi}{\sqrt{\omega_i^2-\sigma^2 k^2_i r_i^2}}.
\end{align}
As $\sigma$ approaches the critical coupling $\sigma_c$, there is a slowing down of the computation due to the divergence of the time-scale $\tau_f \sim (\sigma_c-\sigma)^{-1/2}$. 
As the equation indicates, nodes with high connectivity (i.e hubs) have the longest bit-flip timescales, whereas lower-degree nodes compute on a faster time-scale.

Given this view of the environment as a base computational process, $\bs{b}\mapsto\bs{b}'$, the agent is a control process, and the estimation of the mean field at each node maps to the problem of analyzing a communication channel.
The channel's operational rate is $\beta N$, achieved if the environment has a uniform, i.e. maximum entropy input distribution.
Synchronization changes the input distribution by lowering its entropy---synchronized bits decrease the corpus of words being sent. 
If the agent decodes with high fidelity, then its actions' feedback on the environment is to prevent changes in the input distribution.  
Thus the control process serves to prevent the reduction of the channel's transmission rate.

Putting it together, the background process (environment) transmits at a rate of close to $\beta N$ (bits/s) to the control process (agent). 
Synchronization reduces the transmission rate, to which the control process responds in a manner that counteracts this change.
To address what portion of this information is semantic, we next identify precisely what we mean by viability in this context.

\subsection{Viability Measure $\mathcal{V}$} \label{sec: vb}
As indicated by Fig.~\ref{fig: Order Parameter} the agent-free equilibrium has an expected synchronization greater than the agent-present equilibrium. The effectiveness of the agent at keeping the environment asynchronous, defines its ability to keep the system viable. This provides us with a natural choice within the system dynamics for defining viability in terms of efficacy.

\begin{figure}[t!]
    \centering
    \includegraphics[width=\columnwidth]{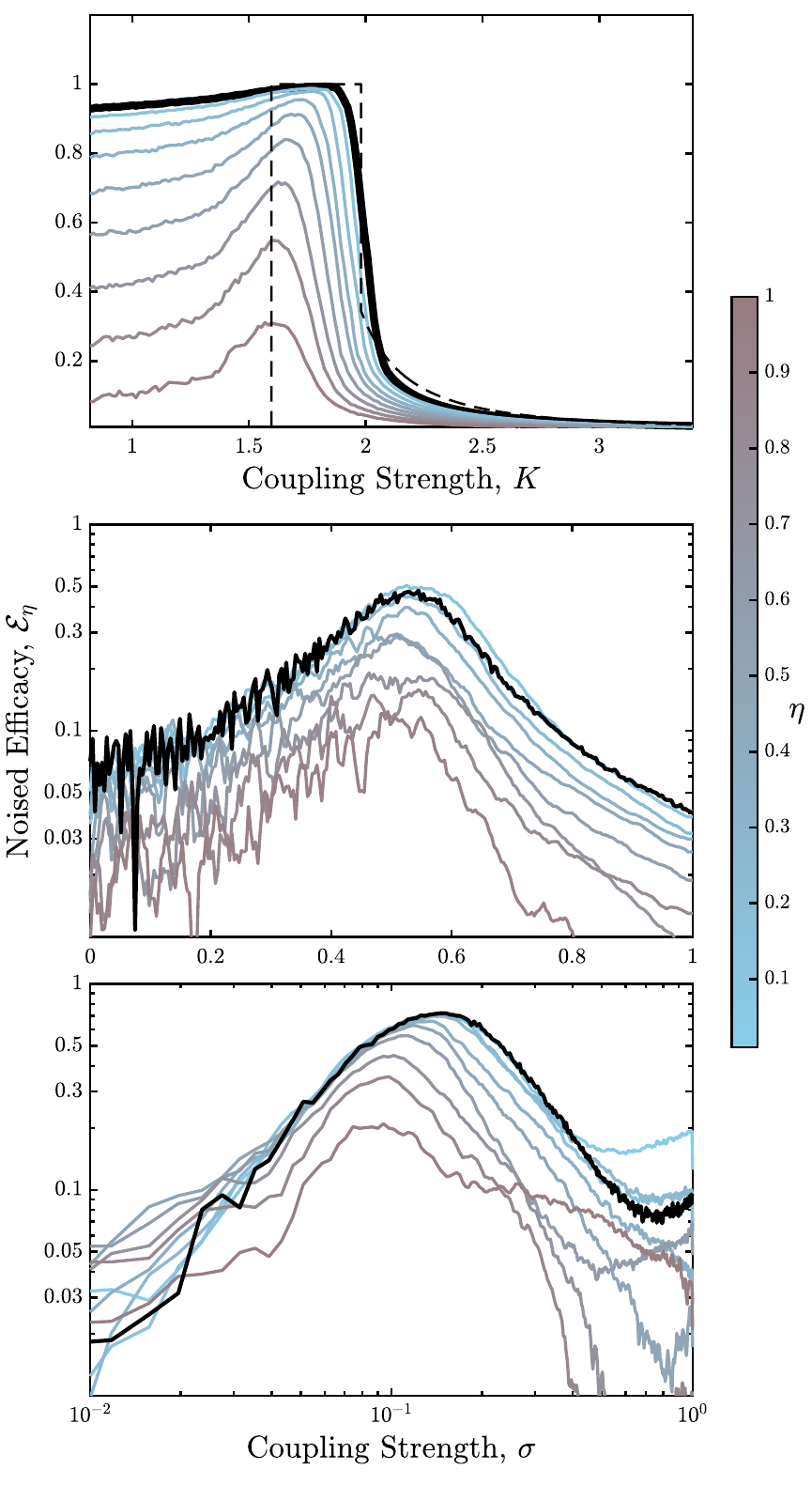}
    \caption{The efficacy, Eq.~\eqref{eq: efficacy} (solid black lines), for agents on fully-connected (top), Poisson (middle), and scale-free graphs (bottom)  graphs with $\alpha =0.5$ and $\beta=0.16$.
    The fully connected curves are averaged over ensembles of $1024$ different graphs; the other two have ensemble sizes of $128$.
    The noised efficacies, Eq.~\eqref{eq: efficacy_eta}, are shown as colored curves with blue corresponding to low noise and pink to high noise. 
    For fully-connected graphs the ensemble is of size $1024$ again, while the Poisson and scale-free graphs have smaller ensembles of $32$.
    The dashed black line in the top panel is our analytical approximation ~\ref{eq: efficacy FC approximation} of the agent efficacy.}
    \label{fig: Noised Efficacy}
\end{figure}

The {\it efficacy} is the relative difference between the expected value of the synchronicity of the environment including the agent, $\mathds{E}[r|\alpha,\beta]$, and without it, $\mathds{E}[r]$, thus,
\begin{align}\label{eq: efficacy}
    \mathcal E =  1-\frac{\mathds E[r|\alpha\beta]}{\mathds E[r]}.
\end{align}
Given that $r$ is a function of $\sigma$, this necessarily implies that the efficacy is also a function of the coupling. 
The more the agent's actions decrease the expected synchronization, the closer the efficacy is to $1$ whereas if the agent is completely ineffectual, the efficacy is $0$.
The {\it viability}, then, is the expected efficacy given the distribution of possible coupling strengths, $\rho(\sigma)$, 
\begin{align}\label{eq:viability}
    \mathcal V &= \mathds{E}_\sigma[\mathcal E]\nonumber\\
    &=\int_0^\infty \!\!\! d\sigma \ \mathcal E(\sigma)\rho(\sigma).
\end{align}
In Fig.~\eqref{fig: Noised Efficacy}, we plot the efficacy (Eq.~\eqref{eq: efficacy} as solid black lines in function of $\sigma$ for the three types of networks. 
In all cases we see a non-monotonic behavior, with a peak in $\mathcal{E}$ at intermediate values of $\sigma$. 
Qualitatively speaking, for low values of the coupling ($\sigma \ll \sigma_c$) the effect of the agent is minimal given that the system is in an asynchronous state and therefore there is no need to act. 
On the other hand in the strong coupling regime $\sigma \gg \sigma_c$, the agent also has low efficacy---given that $\beta \ll 1$, a large fraction of nodes synchronize at a timescale faster than its ability to influence these nodes. 
Thus there is a peak in viability at intermediate values of $\sigma$ where the agent can exert its strongest influence in preventing the synchronization of a non-negligible fraction of oscillators.  

It is instructive to check the fully-connected model, where $\sigma=K/N$, to see if we can make further progress.
In the thermodynamic limit, an exact calculation for the critical coupling for the agent-free environment leads to $K_c = 1.595$, plotted as a blue dashed line in the upper panel of Fig.~\ref{fig: Order Parameter}. 
For the agent-based environment, through a sequence of manipulations, one can arrive at a numerical estimate for the shifted critical coupling $K^A_c\approx 1.988$, plotted as a pink dashed line in the same figure. 
(See Appendices ~\ref{app: fully connected} and ~\ref{app: Agent critical coupling} for details of the calculation, including analytically tractable estimates for $K^A_c$.)

To estimate the efficacy, we note that in environments with coupling below $K_c$, we expect the agent to be ineffective given the environmental dynamics are driven towards asynchronous states.
Above $K_c$ but below $K^A_c$, the efficacy should jump to $1$ since one expects $\mathds{E}_{\alpha\beta}[r]$ to vanish while $\mathds{E}[r]$ does not.
At couplings larger than $K_c^A$ the agent+environment synchronization.
To first order in $\alpha\beta$, the efficacy can be approximated as
\begin{align}\label{eq: efficacy FC approximation}
    \mathcal E \approx\left\{\begin{array}{cc}
        0 & K<K_c \\
        1 & K_c\le K<  K^A_c \\
        \frac{\alpha\beta}{4}\frac{K_c}{K-K_c}& K>  K^A_c.
    \end{array}\right.
\end{align}
This is plotted in the top panel of figure \ref{fig: Noised Efficacy}, where it captures many of the features of the actual efficacy.
We note that our estimate differs in the $K< K_c$ regime.
This is unexpected because as $K\rightarrow 0$ the distribution of phases should approach a Rayleigh distribution describing the magnitude of the sum of a set of randomly directed unit vectors.
In hindsight, this is not hard to understand.
The Rayleigh distribution takes into account even those configurations where the random vectors all point in similar directions.
The agent disrupts these types of arrangements, shifting probability from the tail of the Rayleigh distribution, thereby lowering the expected value of the magnitude i.e. the global order parameter.

With the the fully connected model as reference, similar features are found in the simulations of more complex topologies in Fig.~\ref{fig: Noised Efficacy}; the efficacy is peaked above the critical coupling of the agent-free environment $\sigma_c$, and decreases beyond the critical coupling of the agent-inclusive environment $\sigma^A_c$. 
Based on these results we are now able to both define viability and characterize how it responds to agent dynamics. 
To derive semantic information content in the system we now explore the architecture of correlations in the agent/environment coupling.  

\subsection{Scrambling protocol and semantic content.} \label{sec: spsc}
With the computational description in mind, a natural way of scrambling presents itself.
A noisy channel reduces the information transmission rate, and therefore, the control process will not be able to modulate its behavior in a way that reduces synchronization in the background process, thus affecting its viability. 
Consequently, scrambling the correlations between agent and environment can be implemented by introducing noise in the agent's information about the phase of the neighborhood of node $i$ that it visits.
We interpret this scrambling as a decrease in the fidelity of the sensor used by the  agent to probe the mean field.

Let this noise be parameterized by $\eta\in[0,1]$, where $\eta=0$ corresponds to the agent with $\hat\phi_i=\phi_i$, and $\eta=1$ an agent for whom $\hat\phi_i$ is drawn uniform randomly between $-\pi$ and $\pi$. 
Accordingly, the fidelity of the agents mean field sensor, $1-\eta$, is perfect in the former and is broken in the latter case.
In the range x$0<\eta<1$, the estimator is a random variable $\hat\phi_i\sim \mathcal U(\phi_i-\eta\pi,\phi_i+\eta\pi)$.
Then for node $i$, defining $\eta_c=\min(|\theta_i-\phi_i|/\pi,1-|\theta_i-\phi_i|/\pi)$, the probability that the agent's kick is in the wrong direction is
\begin{align}\label{eq: kick mistake}
    p(\hat \kappa_i\neq\kappa_i|\phi_i,\theta_i) = \left\{
        \begin{array}{ll}
            0 & \text{if }\eta\le\eta_c   \\
            \frac{1}{2}(1-\frac{\eta_c}{\eta}) & \text{if } \eta_c<\eta<1-\eta_c\\
            1-\frac{1}{2\eta} &\text{otherwise.}
        \end{array}
    \right.
\end{align}
For unsynchronized oscillators $|\theta_i-\phi_i|$ takes on all values between $\pm\pi$, so $\eta_c$ will take on values throughout $[0,0.5]$. 
There will be intervals in each period of these oscillators where the agent may make mistakes. 
For synchronized oscillators, $\eta_c = |\sin^{\mhyphen 1}\omega_i/\sigma k_i|/\pi$.
On these, neglecting the back reaction on the network, the agent will continue functioning until $\eta=\eta_c$, at which point, there will be errors in its ability to judge which action to take.

Averaged over all nodes visited by the agent per unit time, the contribution due to noising from both synchronized and unsynchronized oscillators further reduces the information transmission rate from environment to agent.
To quantify the effect of scrambling on the viability we modify Eq.~\eqref{eq: efficacy} by defining the noised efficacy
\begin{align}\label{eq: efficacy_eta}
    \mathcal E_\eta = 1-\frac{\mathds{E}[r|\alpha,\beta,\eta]}{\mathds{E}[r]},
\end{align}
where the expectation in the numerator of the second term is still taken over an ensemble of $\alpha\beta$-agents but ones whose mean field estimators have a degree of noise, $\eta$.
Note that $\mathds{E}[r|\alpha,\beta,\eta=0]=\mathds{E}[r|\alpha,\beta]$, so $\mathcal E_0=\mathcal E$.

We plot the results of introducing noise in Fig.~\eqref{fig: Noised Efficacy} which shows the change in efficacy (colored solid lines) in the range  $0 \leq \eta \leq 1$. For both networks, the peak efficacy is reduced by an order of magnitude for maximal noising; while the overall shape of the curve is preserved with increased noise the curve is increasingly flattened. 
Note that even at maximal noising, where the agent has no correlations with the environment, the efficacy is still non-zero. 
This is easy to understand in terms of the agent acting as a heat bath---even though it is no longer making an informed choice in the way it kicks its local oscillator, the net effect is still to deform the frequency distribution enough to increase the critical coupling to $\sigma^A_c > \sigma_c$, and thus delay the onset of synchronization. 

\begin{figure*}
    \centering
    \includegraphics[width=\textwidth]{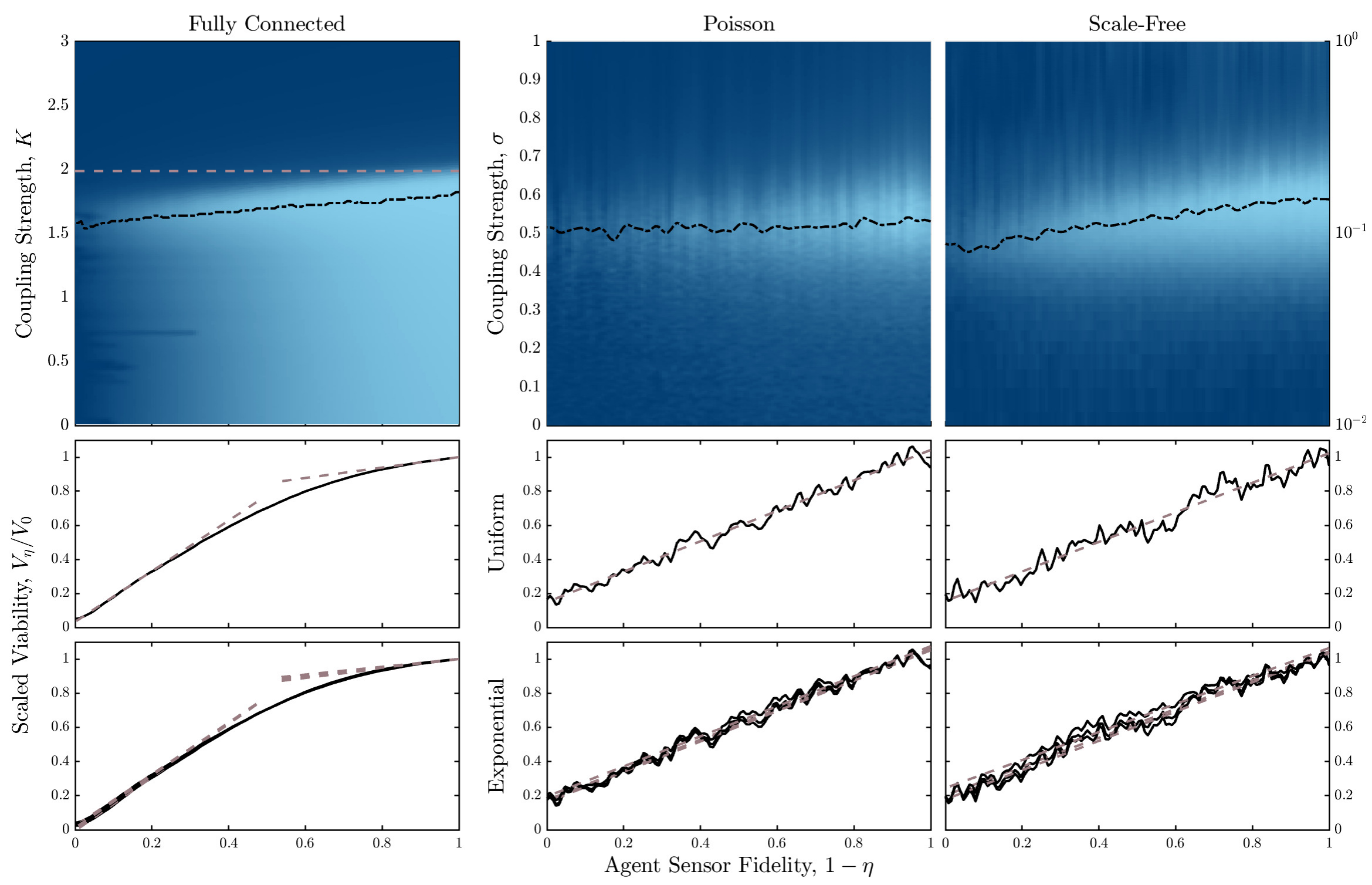}
    \caption{The noised efficacy Eq.~\eqref{eq: efficacy_eta} (top panels) as a function of sensor fidelity (horizontal axis) and coupling strength,$K$, (vertical axis) for fully-connected graphs (left) and $\sigma$ for Poisson graphs (middle), and scale-free graphs (right). 
    For the fully-connected the horizontal dashed pink line is the analytical value of $K_c$.
    In all three panels the dashed black line traces the maximal efficacy at each value of sensor fidelity. 
    The noised viability Eq.~\eqref{eq: viability_eta}, scaled to the actual viability is plotted in the lower two rows of panels as a function of unscrambled information.
    The middle row assumes coupling strengths are drawn from a uniform distribution, while the bottom row has them drawn from three different exponential distributions. 
    The pink dashed lines in these are the best fit linear models, with further details in the text.}
    \label{fig: Viability}
\end{figure*}

As before, the noised efficacy is a function of the coupling strength $\sigma$, so the scrambled viability is the expected noised efficacy over the distribution of possible couplings,
\begin{align}\label{eq: viability_eta}
    \mathcal V_\eta = \int_0^\infty\!\!\!\! d\sigma \ \mathcal E_\eta(\sigma)\rho(\sigma).
\end{align}
In principle, there is no restriction on the distribution $\rho(\sigma)$. 
For illustrative purposes, we use the uniform and exponential distributions.

The efficacy as a function of sensor fidelity (x-axis) and the coupling (y-axis) is shown in the top panels of Fig.~\ref{fig: Viability}, with the light blue regions corresponding to those of high efficacy. 
The fully-connected graph is shown in the left-panel, while the Poisson and scale-free graphs are shown in the middle and right panels respectively.
The width of the regions correlates with sensor fidelity, tapering as the amount of scrambled information increases.
In the fully-connected case we see a strikingly different behavior in the low coupling regime, with the efficacy remaining high up until $\eta\approx 0.5$, as discussed in the last section.
Both Poisson and scale-free graphs do not exhibit this high efficacy at low coupling, which we return to later.
In all cases there is only a slight decrease in peak efficacy (dashed black curves in the top panels) with decreasing sensor fidelity. 
The high-efficacy region occurs for lower coupling strength in the scale-free case, but this is to be expected since $\sigma_c ^{\textrm{scale-free}} < \sigma_c^{\textrm{Poisson}}$ as $\langle k^2 \rangle^{\textrm{scale-free}} > \langle k^2 \rangle^{\textrm{Poisson}}$.

Integrating over all coupling strengths, the viability (scaled to the actual viability), is shown in the middle and bottom panels of Fig.~\ref{fig: Viability}.
In the middle panels a uniform distribution of couplings is used, while in the lower panels three exponential distributions are used.
These differ in their means, chosen as $0.5\sigma_c^A$, $\sigma_c^A$, and $2\sigma^A_c$ with $\sigma_c^A\approx 0.5$.
The solid lines represent simulations, and the dashed lines are linear fits.
The figure indicates that the viability curves are not particularly sensitive to the choice of $\rho(\sigma)$.

The curves for the fully connected graph are best approximated by piece-wise linear fits indicating a transition in the viability per bit (VpB) from $1.5\pm0.05$ at low fidelity to $0.25\pm0.02$ at high fidelity. 
At poor sensor performance each bit of information is meaningful, significantly increasing system viability. This crossover indicates the presence of a semantic threshold. 
Once one crosses the threshold, the fidelity of the sensor becomes less important.
This is analogous to prior work \cite{Sowinski:2023vf}, albeit with a steeper viability plateau.

In both Poisson and scale-free graphs, we no longer see the emergence of a semantic threshold, with a smooth linear dependence of the viability curves with sensor fidelity.
The best linear fits to the curves (plotted as pink dashed lines), imply a constant viability per bit (VpB) in the agent/environment communication channel. 
For Poisson graphs VpB is found to be $0.886\pm0.01$, while for scale-free graphs $0.841\pm0.01$. 
Note that in the absence of noise, the presence of an agent that can accurately estimate the local mean field, has a net effect of lowering the global synchronization  $\approx 20\%$ across the given range of coupling parameters. 
This appears to be independent of the topology, although the decrease in viability with noise is more pronounced in the scale-free case. 
This is likely due to the fact that the agent spends most of its time located in the hubs (given that the probability to visit a node $i$ in a random walk, is $p_i \propto k_i$), and according to Eq.~\eqref{eq: timescale} the computational timescale for hubs is much longer than lower-degree nodes. 

The absence of any plateau, and a constant VpB in the latter two graphs is in stark contrast with the results seen in the fully-connected one and in prior models~\cite{Sowinski:2023vf}. 
There, the viability plateau---marked by a peak in the VpB---differentiated the region of semantic and syntactic information. 
Correlations above the threshold were irrelevant to the forager's viability. 
In these cases, however, every bit of syntactic information is semantic. All correlations the agent has with the environment are essential to the viability of the combined agent-environment system. The fact that the threshold appears in the fully connected case, whereas the qualitative behavior of the viability is indistinguishable in the two sparser graphs, suggests that the key parameter in determining the presence/absence of the threshold is not $\langle k^2\rangle$, but the density of the graph $\langle k \rangle /N$.

\section{Discussion} \label{sec: disc}

The goals of this paper have been twofold. 
First we propose a general road map towards applying semantic information theory (SIT) theory to dynamical systems.  
Building on~\cite{kolchinsky2018semantic,Sowinski:2023vf} we articulate the steps required to cast a system into a form in which one can go beyond syntactic measures of information and consider instead how much of that information contributes to a systems self-maintenance (i.e. viability). 
Broadly, the procedure involves separating the system into agent-environment sub-systems; coarse-graining the DoF and identifying which correlations to follow; identifying an intrinsic viability measure; intervening with a scrambling protocol on the correlations, and measuring the change in viability.  
In this way a system and its dynamics  can be analyzed as agent-environment correlations directly responsible for an agent's effectiveness at maintaining itself, the environment, or both, far-from-equilibrium.  
As we have noted it is the equilibrium state in question that provides the basis for a semantic interpretation of this information.  
Thus determination of such equilibria becomes an essential step in the applying SIT to a system.  
In some cases such equilibria may be obvious, such as the death of an agent~\cite{Sowinski:2023vf}, however, as we have shown in this paper it is possible that the equilibria of interest, and hence proper viability functions, may require more effort to determine.

Our second goal was to operationalize the road map by exploring a specific model to demonstrate how SIT can bring new insights to previously studied problems.  
Loosely motivated by biological models such as the neurodynamics of epilepsy and other biologically pathological states, we introduced an agent-based variant of the well-known Kuramoto model. 
Treating the original model on an arbitrary network as an environment, with an exogenously dependent coupling, an agent is introduced tasked with preventing/delaying the onset of synchronization when the environmental coupling exceeds the critical value. 
The modified system is a stochastic version of the original model where the effect of the agent is to function as a heat bath that deforms the distribution of frequencies for the oscillators, thus delaying the onset of synchronization. 
Articulating the resulting correlational structure for agent and environment (the oscillators), allows us to cast the model from a computational (information theoretic) perspective, where the agent-environment dynamics can be analyzed as a communication channel. 
Adding noise to this channel enables the computation of the system's viability as a function of the level of scrambling of the channel and the viability per bit.  
As in previous explorations of semantic information~\cite{Sowinski:2023vf}, our results showed a semantic threshold in fully-connected graphs.
Both Poisson and scale-free graphs, however, lacked this feature.
The results indicate that semantic information is affected by network topology, with the relevant parameter being the density of the graph as opposed to the degree distribution. 
For sparser graphs, every bit of information in the communication channel between the agent and environment is essential for the system's viability, whereas dense graphs display a semantic threshold.

In order for an agent to act on information about its environment, it must be able measure this information via some kind of a sensor. 
The presence or absence of a semantic threshold in a system is then related to the fidelity of the sensor and how it is optimized with respect to the environment. 
The difference between the forager model and the control-feedback Kuramoto model may well be due to the fact that the sensor in the former case is over-optimized with respect to its environment, whereas in the latter case it senses \emph{only} the information it needs to maintain viability. 
In biological systems, evolutionary adaptations have led to sensors in organisms with varying amounts of fidelity~\cite{Orr:2005ng}. 
For instance, mice, which are nocturnal creatures, cannot see particularly well at night, but have evolved to adopt large ears that give them excellent hearing~\cite{Morsli:1998va}. 
Therefore, explorations of SIT on real biological systems may or may not display a semantic threshold depending on the organism, its evolutionary history, and the context in which a viability function is defined. 
The connection between sensor fidelity and semantic information will be the subject of future explorations.

\section{Acknowledgements}
The authors thank Artemy Kolchinsky for his comments on the manuscript, and the Center for Integrated Research Computing (CIRC) at the University of Rochester for providing computational resources and technical support.
A heartfelt thanks to the thorough readings and comments made by the anonymous reviewers for which we credit our discovery of the semantic threshold in the fully-connected model.
This project was partly made possible through the support of Grant 62417 from the John Templeton Foundation. 
The opinions expressed in this publication are those of the author(s) and do not necessarily reflect the views of the John Templeton Foundation.

\appendix

\section{Simulation methodology}\label{app: simulation methodology}
To acquire the data discussed in this paper we performed simulations written in MATLAB and run on the Bluehive cluster at the University of Rochester. 
For simulations with and without agents, the underlying graphs were either fully-connected, Poisson, or scale-free.
The latter were constructed via the configuration model~\cite{Newman_nets}, resulting in graphs with $N\approx 1000$ nodes and average degree of  $\langle k\rangle\approx 4$.

For the Poisson and scale-free environments we constructed $1024$ distinct graphs for each of the $256$ values of the coupling, $\sigma\in[0,1]$.
The same was done for the full-connected environments, except that the coupling was in the range $K\in[0,4]$.
The environments were initialized with random phases drawn uniformly, and random natural frequencies drawn from a Gaussian distribution with unit standard deviation.
Using $\Delta t=1/16$, each environment was run for $1500$ timesteps to ensure the decay of any transients; data was then collected for the following $512$ timesteps and averaged to give the synchronization of each environment.

With the introduction of the agent, $\alpha = 0.5$ $\beta = 0.16$, the number of distinct Poisson and scale-free graphs was reduced to $32$, leaving the other numbers mentioned above the same. 
Furthermore, the experiments were now done for an additional $128$ values of the scrambling parameter, $\eta$, equally spaced in the range $[0,1]$.

\section{Derivation of Continuum Dynamics}\label{app: continuum approximation}
\begin{figure}
    \centering
    \includegraphics[width=0.95\columnwidth]{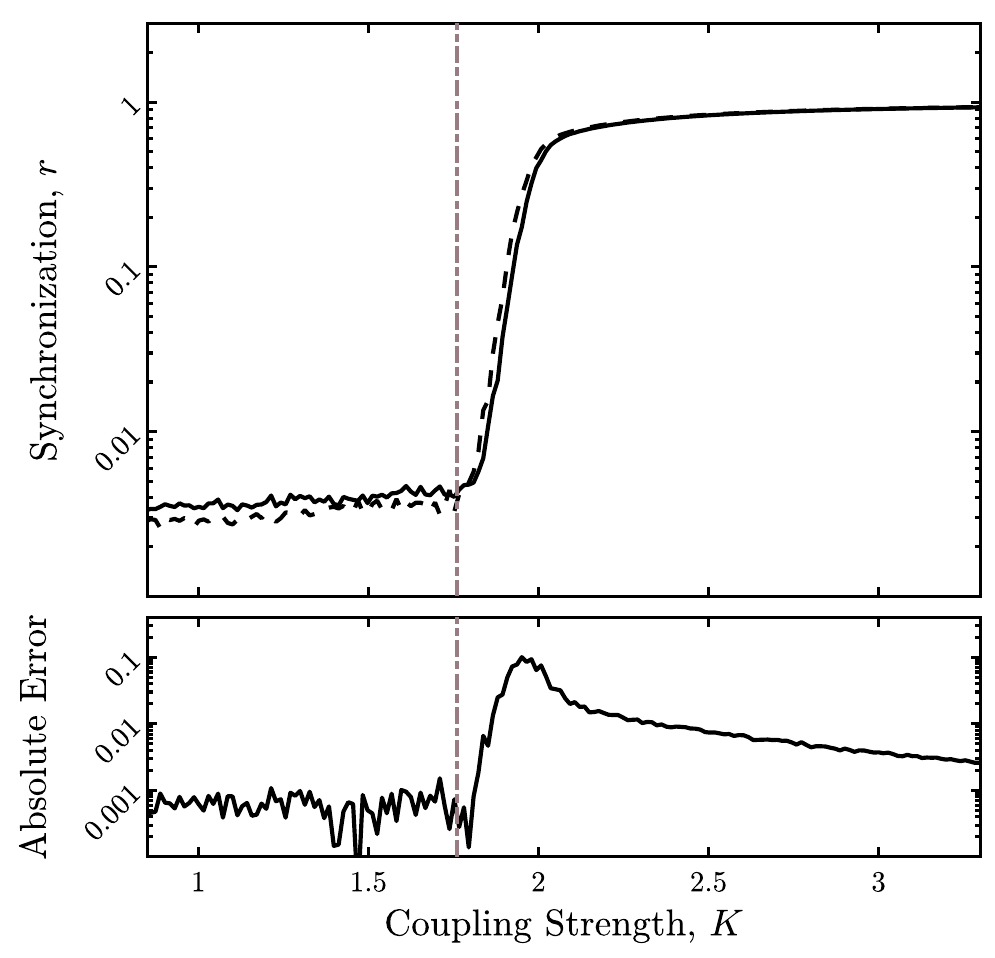}
    \caption{In the top panel we have a comparison of the order parameter in the the agent-based model (solid curve) and the stochastic differential equation eq.\ref{eq: Langevin EoM} (dashed curve). The bottom panel shows the absolute error between the two. Note that the y-axis is logarithmic in order to exaggerate the underestimation of the SDE.}
    \label{apfig: continuum comparison}
\end{figure}

The agent moves between nodes on a timescale $\tau$.
During a time interval $\Delta t$, the agent makes $J= \Delta t/\tau = \beta N\Delta t$ jumps between nodes.
Due to the topology of the environment the agent visits nodes with higher degree more often than those with lesser degree.
The probability of visiting a node $i$ is proportional to its degree, $p_i\propto k_i$ (with normalization $N\langle k\rangle$).
Each visit results in a perturbation of strength $\alpha$, so the expected perturbation to that node is $\Delta \theta_i = \alpha p_i J=\alpha\kappa_i\beta k_i\Delta t/\langle k\rangle$,
Since visits are rare, the perturbations are a Poisson arrival process.
On timescales much larger than a single jump, the expected number of visits is high enough to model the Poisson process as a fixed rate with Gaussian fluctuations.
Taken together, this means that we can think of the agent's actions on each oscillator as approximated by a continuous perturbation of strength
\begin{align}
    \left.\frac{d\theta_i}{dt}\right|_\text{cont.} = \alpha\beta\kappa_i\frac{k_i}{\langle k\rangle}
\end{align}
modulated by, heuristically, a stochastic fluctuation of size
\begin{align}
    \left.\frac{d\theta_i}{dt}\right|_\text{stoch.} = \sqrt{\alpha\beta\frac{k_i}{\langle k\rangle}}\frac{1}{\sqrt{\Delta t}}
\end{align}
Note the presence of the $\sqrt{\Delta t}$ above, which motivates our quantification of this heuristic by the replacement of this term with a Wiener process.
To see how well the continuum model works we plot it against the agent based model in figure \ref{apfig: continuum comparison}.

\section{Review of Fully Connected Model}\label{app: fully connected}
In this appendix we review the calculation of the critical coupling in the original fully connected model.
There we have that $\sigma = K/N$; the coupling needs to scale inversely with the number of nodes because in a fully-connected graph the number of connections, by definition, scales with $N$.
Our starting point is therefore Eq. ~\eqref{eq: Environment EoM 1}, the mean field equation of motion.
We work under the simplifying assumption of the long-time, thermodynamic limit, $t\rightarrow\infty, N\rightarrow\infty$, so that fluctuations in the synchronization order parameter, $r$, vanish, allowing us to treat it as constant.
In what follows we will derive a self-consistency constraint on the synchronization, from which everything else will follow. 

\begin{figure}
    \centering
    \includegraphics[width=\columnwidth]{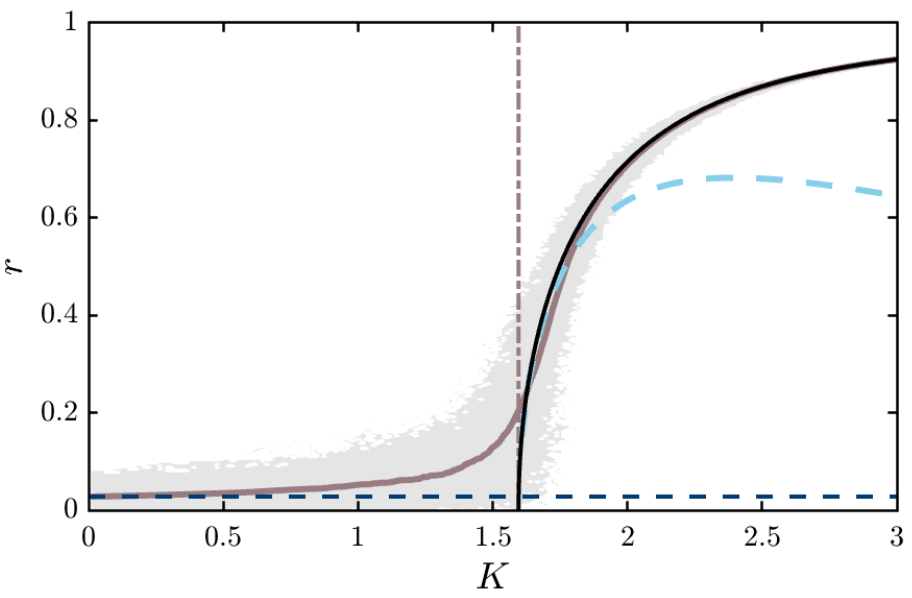}
    \caption{Synchronization in the fully connected Kuramoto model with a Gaussian distribution of natural frequencies. The vertical dashed-dot line is at the critical coupling, $K_c = 2\sqrt{2/\pi}$. The dashed horizontal line is the Rayleigh distribution mean. The light grey region is data from our simulations, with the mean a solid curve going through it. The black curve is the analytical solution eq.~\ref{eq: critical solution gaussian}. The dashed blue curve is the second order approximation to that solution.}
    \label{apfig: phase diagram}
\end{figure}

Consider an oscillator with natural frequency $\omega$.
If $|\omega|>Kr$, then $\dot{\theta}$ never vanishes, and the oscillator is unsynchronized. 
On the other hand if $|\omega|\le Kr$, the oscillator synchronizes, approaching a phase of $\theta\rightarrow \phi +\sin^{-1}\omega/Kr$.
We use these facts to divide the set of oscillator indices, $\mathcal I$, into synchronized, $\mathcal I_s=\{i:|\omega_i|<Kr\}$, and unsynchronized, $\mathcal I_u=\mathcal I\backslash \mathcal I_s$, subsets.
The sum in the definition of the order parameter becomes
\begin{align}
    re^{i\phi} &= \frac{1}{N}\sum_{i\in\mathcal I} e^{i\theta_i}\nonumber\\
    &=\frac{1}{N}\sum_{i\in \mathcal I_u} e^{i\theta_i} + \frac{1}{N}\sum_{i\in\mathcal I_s} e^{i\theta_i}\nonumber
\end{align}
The thermodynamic limit assumption allows us to disregard the first term - the unsynchronized oscillators will be distributed uniformly across phases, cancelling.
For the second term, we use the long-time assumption to set the phases to their synchronized values noted above, getting
\begin{align}
    r=\frac{1}{N}\sum_{i\in\mathcal I_s}e^{i\sin^{-1}\frac{\omega_i}{Kr}}\nonumber
\end{align}
Then we insert an integral over frequencies containing the delta function $\delta(\omega-\omega_i)$.
The sum gives a density of states, which in this case is the frequency distribution and a Jacobian factor in going from oscillator index to oscillator frequency, resulting in
\begin{align}
    r = \int_{-Kr}^{Kr}\!\!\! d\omega \ g(\omega)\left(\sqrt{1-\frac{\omega^2}{K^2r^2}} + i\frac{\omega}{Kr}\right).\nonumber
\end{align}
Since we work with symmetric frequency distributions, the imaginary piece of the above vanishes.
In what remains, we perform the substitution $\omega = K r x$, to get the self-consistency constraint
\begin{align}\label{eq: consistency constraint}
    \frac{1}{K} = 2\int_{0}^{1}\!\!\!\!\! dx \ \sqrt{1-x^2} g(Krx)
\end{align}
In the case of a standardized Gaussian, the integral can be performed leading to an algebraic self-consistency constraint,
\begin{align}\label{eq: critical solution gaussian}
    \frac{1}{K}=\sqrt{\frac{\pi}{8}}e^{-\frac{K^2r^2}{4}}(I_0(\frac{K^2r^2}{4})+I_1(\frac{K^2r^2}{4})),
\end{align}
where the $I_\nu$ are modified Bessel functions.
This transcendental equation is plotted on the $(K,r)$ plane, showing the well-known synchronization phase transition in figure \ref{apfig: phase diagram}.

For more general symmetric distributions, proximity to the critical point is still amenable to analysis.
When $K>K_c$, $r>0$, and as $K\rightarrow K_c$ from above, $r\rightarrow 0$.
This observation motivates an expansion in $Kr$ for the frequency distribution
\begin{align}
    g(Krx) = \sum_{n=0}^\infty \frac{1}{2n!}g^{(2n)}(0)(Kr)^{2n}x^{2n},
\end{align}
where the even symmetry of $g$ has been used.
Plugging into the self-consistency constraint results in integrals easily recognizable as Euler beta functions. Rewriting this in terms of Euler gamma function identities results in
\begin{align}
    \frac{1}{K}=\frac{\pi}{2}\sum_{n=0}^\infty \frac{g^{(2n)}(0)K^{2n}r^{2n}}{4^n n!(n+1)!}.
\end{align}
Setting $K=K_c$ so that $r=0$ gives the critical coupling in terms of the center of the frequency distribution,
\begin{align}\label{eq: critical coupling}
    \frac{1}{K_c} = \frac{\pi}{2}g(0).
\end{align}
If the distribution is monotonically decreasing for $\omega>0$, and does not have vanishing curvature at $\omega=0$, then we can go to second order to extract the behavior near the critical coupling.
Analysis of
\begin{align}
    \frac{1}{K}&=\frac{1}{K_c}-\frac{\pi}{16}|g''(0)|K^2r^2\nonumber\\
    &\Downarrow\nonumber\\ \label{eq: crit behavior}
   r^2 &=  8\frac{g(0)}{ |g''(0)|}\frac{K-K_c}{K^3}
\end{align}
gives a critical exponent of $1/2$.

\section{Critical Coupling with an Agent}
\label{app: Agent critical coupling}
Here we present several methods for estimating the critical coupling in the presence of an agent. 
The first of utilizes a Gaussian approximation for the shape of the effective distribution, whereas the next two use a parameterized ansatz for the effective distribution.

\subsection{Gaussian Approximation}

Here we approximate the effective distribution as a Gaussian, showing that the increase in variance when $g(0)$ is replaced with $g^A(0)$ in equation~\ref{eq: critical coupling} results in a $K_c^A>K_c$.
As described at the end of section \ref{sec: agent-effect}, the frequency distribution is first cut at $\omega=0$, and the positive (negative) half shifted by positive (negative)$\alpha\beta$.
When this is done for a standardized Gaussian, a quick calculation finds the mean unchanged, and the variance now $1+2\sqrt{2/\pi}\alpha\beta+\alpha^2\beta^2$.
Next, the convolution with a heat kernel of variance $\alpha\beta$, results in the variance of the effective distribution,
\begin{align}
    \langle \omega^2\rangle_A&=1+(1+2\sqrt{\frac{2}{\pi}})\alpha\beta + \alpha^2\beta^2\\
    &\ge 1.\nonumber
\end{align}

We next assume the effective distribution is also a Gaussian, with the above variance.
The peak of the effective distribution related to the original is
\begin{align}
    g^A(0)=\frac{g(0)}{\langle\omega^2\rangle_A^{1/2}}
\end{align}
The critical coupling is consequently shifted according to,
\begin{align}
    K_c\rightarrow K^A_c &=  \langle\omega^2\rangle_A^{1/2}K_c.
\end{align}
In our simulations, $\alpha\beta = 0.08$, this corresponds to $ K^A_c\approx 1.75829$. This however is an
underestimate of the true critical coupling. 

\subsection{Parameterized Effective Distribution}
\begin{figure}[h]
    \centering
    \includegraphics[width=\columnwidth]{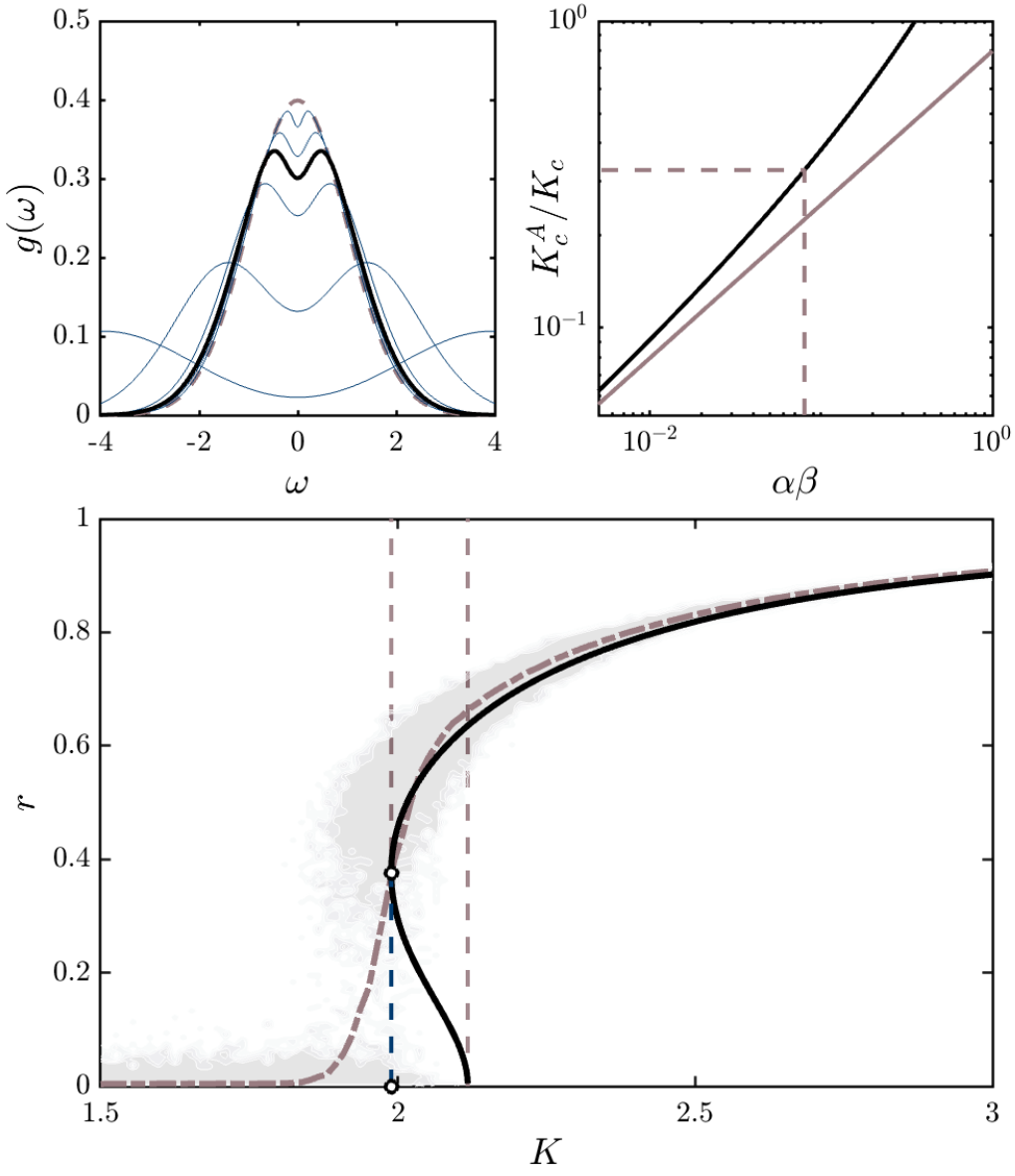}
    \caption{In the top left panel we have the original Gaussian frequency distribution, $g(\omega)$, plotted with a dashed line. The thick black curve is the effective distribution, $g^A(\omega)$, while the thin blue lines are representative of smaller and higher $\alpha\beta$.
    The top right panel is the behavior of the critical coupling as a function of $\alpha\beta$; the full non-linear form shown in black, and the linearized form in pink. The dashed line is the corresponding estimate for $K_c ^A$.
    The bottom panel shows the full analysis of the consistency constraint. Details are in the text.}
    \label{apfig: numerics}
\end{figure}

In order for an improved estimate, we next calculate the critical coupling from the parameterized effective distribution rather than approximating it as a Gaussian.
The initial cut in the distribution and shift of frequencies results in 
\begin{align}
    g(\omega)\mapsto\tilde{g}(\omega)&=\Theta(-\alpha\beta-\omega)g(\omega+\alpha\beta)\nonumber\\
    &\hspace{10pt}+\Theta(\omega-\alpha\beta)g(\omega-\alpha\beta),
\end{align}
where $\Theta$ is the Heaviside step function.
It is easily seen that the first term is the negative part of the distribution shifted to the left, and the second is the positive part shifted to the right.
Convolving with the heat kernel, $h_{\alpha\beta}$, results in
\begin{widetext}
\begin{align}
    g^A(\omega) &= \frac{1}{\sqrt{2\pi\alpha\beta}}\int_{-\infty}^\infty \!\!\! d\omega'\ \tilde{g}(\omega-\omega')e^{-\frac{\omega'^2}{2\alpha\beta}}\\
    &=\frac{1}{2\pi\sqrt{\alpha\beta}}\left[
    \int_{\omega\!+\!\alpha\beta}^\infty \hspace{-20pt} d\omega' \exp\left(-\frac{1}{2}(\omega'\!-\!(\omega\!+\!\alpha\beta))^2\!-\!\frac{1}{2}\frac{\omega'^2}{\alpha\beta}\right)
    \!+\!
    \int^{\omega-\alpha\beta}_{-\infty} \hspace{-23pt} d\omega'\exp\left(-\frac{1}{2}(\omega'\!-\!(\omega\!-\!\alpha\beta))^2\!-\!\frac{1}{2}\frac{\omega'^2}{\alpha\beta}\right)
    \right]\\
    &=\frac{1}{2\sqrt{2\pi}\sqrt{1+\alpha\beta}}
    \left(
    e^{-\frac{1}{2}\frac{(\omega+\alpha\beta)^2}{1+\alpha\beta}}
    \text{erfc} 
    \left [ \frac{\omega+\alpha\beta}{\sqrt{2\alpha\beta(1+\alpha\beta)}}
    \right]
    +
    e^{-\frac{1}{2}\frac{(\omega-\alpha\beta)^2}{1+\alpha\beta}}\text{erfc} \left[\frac{-\omega+\alpha\beta}{\sqrt{2\alpha\beta(1+\alpha\beta)}}\right]
    \right),
\end{align}
\end{widetext}
where erfc$(x)$ is the complementary error function. This suggests a bimodal form for 
$g^A(\omega)$ constructed of two overlapping Gaussian distributions; the smaller $\alpha\beta$ is, the greater the overlap, giving the appearance of a Gaussian with a dimple at its center.
Several of these effective distributions are plotted in the upper left panel of figure~\ref{apfig: numerics}.
The solid black curve represents simulation data ($\alpha\beta=0.08$), and the thin blue curves are shown for a range $\alpha\beta$ above and below 0.08. 
For comparison the original Gaussian distribution, $g(\omega)$, is plotted as a dashed line.

Plugging in $\omega=0$ into the effective distribution, and then inserting the result into equation~\ref{eq: critical coupling}, shows how $\alpha\beta$ parameterizes the change in the critical coupling
\begin{align}
    K_c^A&=\sqrt{1+\alpha\beta}\frac{e^{\frac{1}{2}\frac{(\alpha\beta)^2}{1+\alpha\beta}}}{\text{erfc}\left[\sqrt{\frac{\alpha\beta}{2(1+\alpha\beta)}}\right]}K_c\\
    &\approx (1+\sqrt{\frac{2}{\pi}\alpha\beta}+\frac{\pi+4}{2\pi}\alpha\beta+\cdots)K_c.
\end{align}
In the second line we have expanded for $\alpha\beta\ll1$; we see that the critical coupling is a Taylor series in $\sqrt{\alpha\beta}$.
Furthermore, computing the second derivative of the effective potential gives a critical exponent of $1/2$ for the behavior of the new critical coupling; in the Gaussian approximation this critical exponent is $1$.
This behavior of $K_c^A/K_c$ is plotted in the upper right panel of figure~\ref{apfig: numerics}.
The solid black is the full non-linear behavior, while the straight pink line is the linearized behavior (quadratic in $\sqrt{\alpha\beta}$). 
The dashed line is at $\alpha\beta=0.08$ giving us a second estimate for the critical coupling, $K_c^A \approx 2.11751$.

This estimate turns out to be a bit high. 
Getting an even better estimate of the critical point is difficult analytically, as the consistency constraint, to the best of our knowledge, cannot be integrated by hand.
Numerical integration, on the other hand, can be quickly implemented to find $r=r(K)$, which is plotted in the bottom panel of figure~\ref{apfig: numerics} with the solid black curve.
The order parameter is multi-valued in the range $K\in [1.9887,2.118]$, where the right boundary coincides with the previous estimate.
Plotting this alongside the simulation data (the grey region) shows us the estimate is high and that the left end of the range is a better estimate of the critical coupling.
Both of these are plotted as the vertical dashed lines in the panel.
The dashed pink curve is the expected value of the synchronization, which matches very well with numerical simulations. In the thermodynamic limit the synchronization curve has a discontinuity at $K_c^A\approx 1.9887$, and drops to $0$ at smaller values of the coupling.

The plot of the consistency constraint of the effective distribution in figure~\ref{apfig: numerics} matches the major features of the data.
This gives us reasonable confidence that the above analysis using the method of an effective distribution is a valid way of describing the agent-environment system.
That said, the match is not perfect, as one can see from the slight underestimation of the data in the figure.
Since the ensemble size is large, $N=1024$, we do not think that this is due to not being at the thermodynamic limit, allowing us to conjecture that there is yet some refinement to this method which we leave for future work.
\end{document}